\documentclass[prl,twocolumn,showpacs,preprintnumbers,amsmath,amssymb]{revtex4}
\usepackage{graphicx}
\usepackage{dcolumn}
\usepackage{bm}
\usepackage{mathrsfs}
\usepackage{subfigure}
\usepackage{natbib}

\begin{document}

\title{Spatially-dependent sensitivity of superconducting meanders\\ as single-photon detectors}

\author{G. R. Berdiyorov}
\affiliation{Departement Fysica, Universiteit Antwerpen,
Groenenborgerlaan 171, B-2020 Antwerpen, Belgium}
\author{M. V. Milo\v{s}evi\'{c}}
\affiliation{Departement Fysica, Universiteit Antwerpen, Groenenborgerlaan 171, B-2020 Antwerpen, Belgium}
\author{F. M. Peeters}
\email{francois.peeters@ua.ac.be}
\affiliation{Departement Fysica, Universiteit Antwerpen,
Groenenborgerlaan 171, B-2020 Antwerpen, Belgium}

\date{\today}

\begin{abstract}

The photo-response of a thin current-carrying superconducting stripe
with a 90-degree turn is studied within the time-dependent
Ginzburg-Landau theory. We show that the photon acting near the
inner corner (where the current density is maximal due to the
current crowding [J. R. Clem and K. K. Berggren, Phys. Rev. B {\bf 84}, 174510
(2011)]) triggers the nucleation of superconducting vortices at
currents much smaller than the expected critical one, but {\it does
not} bring the system to a higher resistive state and thus remains
undetected. The transition to the resistive state occurs only when
the photon hits the stripe away from the corner due to there uniform
current distribution across the sample, and dissipation is due to
the nucleation of a kinematic vortex-antivortex pair near the photon
incidence. We propose strategies to account for this problem in the
measurements.
\end{abstract}

\pacs{74.78.Na, 73.23.-b, 74.25.Fy}

\maketitle

Superconducting current-carrying thin-film stripes have recently received a revival of interest due to their promising
application for single-photon detection \cite{Anant} with a high maximum count rate, broadband sensitivity, fast
response time and low dark counts \cite{Goltsman,Kerman,Marsili}. The single-photon absorption event leads to the
formation of a nonequilibrium ``hotspot'' with suppressed superconductivity, the area of which grows in time, forming a
normal belt across the strip \cite{Kadin}. The latter leads to redistribution of the current between the now resistive
superconductor and a parallel shunt resistor, where a voltage pulse is detected, before a hotspot region cools down (on
a timescale of few hundred picoseconds \cite{semenovJMO}) and the system returns to its initial superconducting state.
Although the hotspot mechanism nicely explains the photon detection in the visible and near UV range, the detection
mechanism in the near-infrared range is still debated (see e.g. Ref. \cite{hofherr} and references therein).
Superconducting fluctuations, e.g., excitation of superconducting vortices \cite{semenovSST,Engel,bartolf,bula,zotova},
have been put forward as an explanation. Dissipative crossing of such vortices, which hop over the edge barrier, or are
created due to the unbinding of thermally activated vortex-antivortex pairs, provides a good description of the
experiment \cite{bartolf}. This vortex-based mechanism was also shown to be the dominating origin for dark counts
\cite{bula,zotova,Yamashita}, which leads to decoherence in the photon detection process. 

To increase the efficiency of the photon counting, detectors are usually fabricated as a meandering superconducting
wire \cite{Kerman}. However, these structures are vulnerable to edge imperfections \cite{Kerman}, which significantly
reduce the photon counting rate. Moreover, the critical current in these systems is mostly determined by sharp inner
corners where the supercurrent density is maximal due to current crowding \cite{clem,clem2}. While the appearance of
edge imperfections can be reduced by present day technology \cite{Kerman}, the effect of current crowding in such
meandering geometries still demands further investigations. In this work we therefore study the effect of the turns of
a meandering superconducting stripe on the response to a single-photon absorption event. Counterintuitive to many, we
reveal that the current crowding at meandering turns does not facilitate dissipative vortex crossings upon the photon
impact. Actually, the turning corner is virtually insensitive to photon absorption, which must be accounted for in
practice.

\begin{figure}[b]
\includegraphics[scale=0.45]{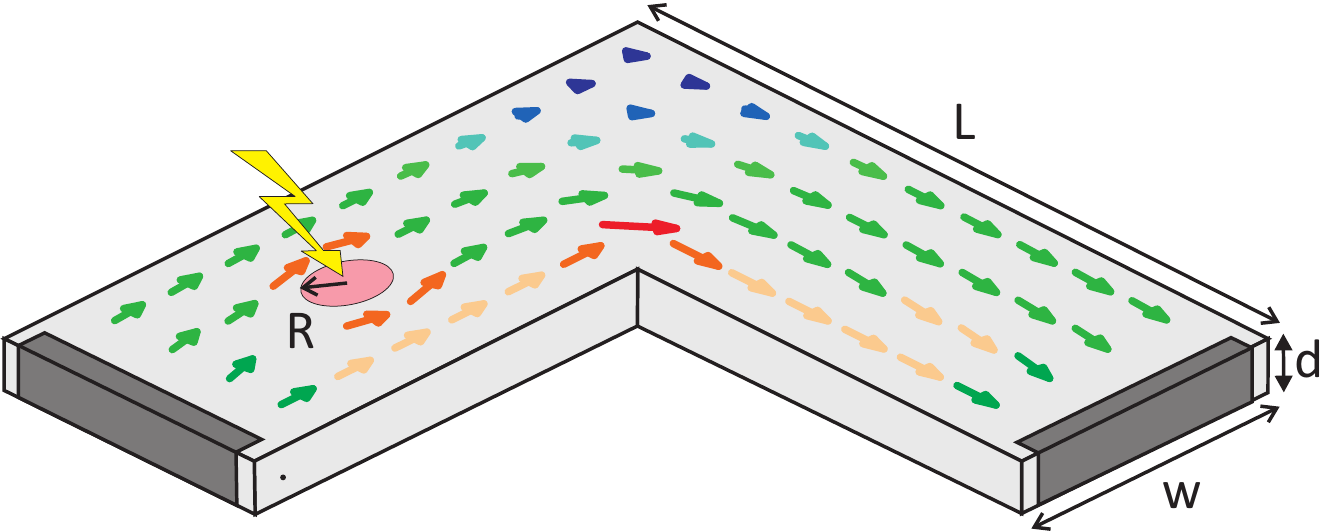}
\caption{\label{fig1} (Color online) The model system: a superconducting stripe (width $w$, length $L$ and thickness
$d$$\ll$$\lambda, \xi$) with a 90$^{\circ}$ turn. The current is applied via normal-metal contacts and the voltage is
measured at a small distance away from these leads. Impact of a photon is modeled by a hotspot with radius $R$. Arrows
indicate the supercurrent distribution.}
\end{figure}

We consider a superconducting strip (with thickness $d\ll\xi,\lambda$ and width $w\ll\Lambda=2\lambda^2/d$, where
$\xi,\lambda$ are the coherence length and magnetic penetration depth) with a 90-degree turn in the presence of a
transport current (see Fig. \ref{fig1}). For this system we solve the following generalized time-dependent
Ginzburg-Landau equation \cite{kramer}
\begin{eqnarray}
\frac{u}{\sqrt{1+\gamma^2|\psi|^2}}\left(\frac{\partial}{\partial t}+i\varphi
+\frac{\gamma^2}{2}\frac{\partial |\psi|^2}{\partial t} \right) \psi =(\nabla-i\mathbf{A})^2\psi\nonumber\\
+(1-T-|\psi|^2)\psi,
\end{eqnarray}
which is coupled with the equation for the electrostatic potential $\Delta\varphi={\rm
div}\{\textrm{Im}[\psi^*(\nabla-{\rm i}{\bf A})\psi]\}$. Here distance is scaled to $\xi(0)$, the vector potential
${\bf A}$ is in units of $\Phi_0/2\pi\xi(0)$, temperature $T$ is in units of $T_c$, time is in units of
$t_0=4\pi\lambda(0)^2/\rho_n c^2$ ($\rho_n$ is the normal state resistivity), and voltage is scaled to
$\varphi_0=\hbar/2e t_0$.
Using $\rho_n$$=$$18.7$ $\mu\Omega$cm, $\xi(0)=4.2$ nm and $\lambda(0)=390$ nm,
which are typical for NbN thin films \cite{semenovSST}, one obtains
$t_0$$\approx$$1.25$ ps and $\varphi_0$$\approx$$0.26$ mV at
$T=0.9T_c$, which will be the working temperature in our
simulations. The coefficients $u$ and
$\gamma$ are chosen as $u=5.79$ and $\gamma=10$ \cite{kramer}. To
model the thermal coupling of our sample to the substrate and the
change of the local temperature we use
the heat diffusion equation \cite{thermal}: 
\begin{eqnarray}
\nu \frac{\partial T}{\partial t}=\zeta\Delta T+j_n^2-\eta(T-T_0),
\end{eqnarray}
where $T_0$ is the bath temperature, $\nu$ is the effective heat
capacity, $\zeta$ is the effective heat conductivity coefficient,
$j_n$ is the normal current density, and $\eta$ is the heat transfer
coefficient which governs the heat removal from the sample.
Following the approach of Ref. \cite{denis} we used $\nu=0.05$,
$\zeta=0.05$ and $\eta=2\cdot10^{-3}$. This allows us to treat the
formation and expansion of photon-induced hotspots in our approach.
We solve the above equations self-consistently on a 2D Cartesian
grid using the Euler and multi-grid iterative procedures. We use
superconducting-vacuum boundary conditions $(\nabla-{\rm i}{\bf
A})\psi|_n=0$, $\nabla\varphi|_n=0$ and $\nabla T|_n=0$ at all
sample boundaries, except at the current contacts where we use
$\psi=0$, $T=T_0$ and $\nabla\varphi|_n=-j$, with $j$ being the
applied current density in units of
$j_0=c\phi_0/8\pi^2\xi(0)\lambda(0)^2$.

\begin{figure} [t]
\includegraphics[width=\linewidth]{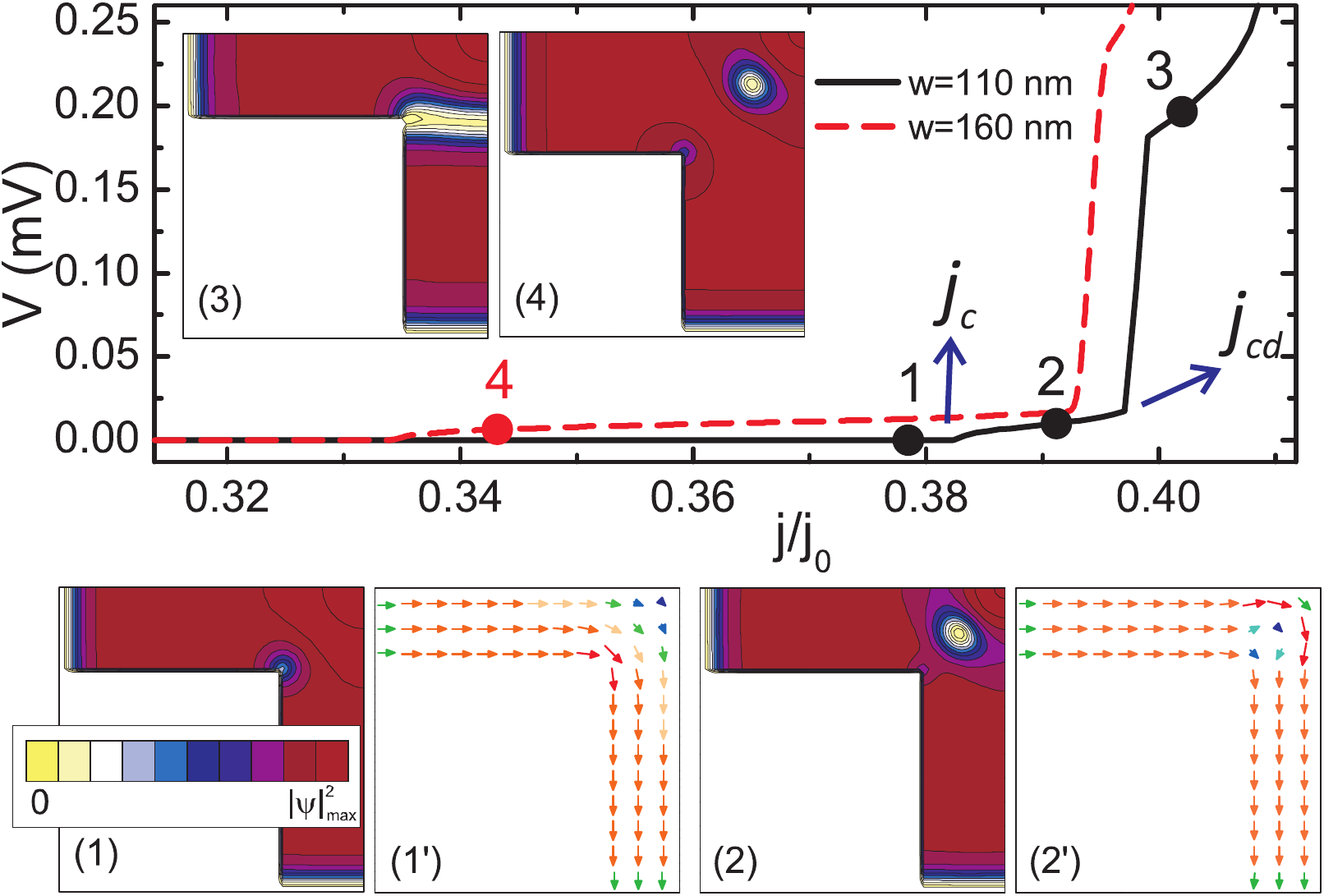}
\caption{\label{fig2}(Color online) $I-V$ characteristics of the
sample of size $L=400$ nm and width $w=110$ nm (solid black curve)
and $w=160$ nm (dashed red curve) at $T=0.9T_c$. Panels (1-4) show
the snapshots of the Cooper-pair density and panels ($1'$, $2'$) the
supercurrent density for the values of the applied current indicated
in the main panel.}
\end{figure}

As a representative example, we consider a superconducting stripe as shown in Fig. \ref{fig1}, of length $L=400$ nm and
width $w=110$ nm, where we use the parameters for NbN thin films (i.e., $\xi(0)=4.2$ nm and $\lambda(0)=390$ nm). Time
averaged voltage vs. applied current characteristics of the sample is shown in Fig. \ref{fig2} for two values of the
sample width $w$. With increasing the external current, zero resistance of the sample is maintained up to a threshold
current density $j_c=0.382j_0$ (see solid black curve), above which the system goes into the resistive state. This
resistive state is characterized by the periodic nucleation of vortices near the inner corner (see panels 2 and 2$'$),
where the current density is maximal due to the current crowding \cite{clem,clem2} (see panels 1 and 1$'$). However,
the distribution of the supercurrent is strongly inhomogeneous, and decays fast away from the inner corner towards the
outer one. The Lorentz force drives the nucleated vortex towards the outer corner of the sample where it leaves the
sample (panel 2), but this motion is slow and weakly dissipative. The nucleation rate of vortices in the inner corner
increases with further increasing the applied current, and at sufficiently large current (labeled $j_{cd}$) the system
transits to a higher dissipative state with a larger voltage jump, characterized by fast-moving (kinematic) vortices
(see panel 3) \cite{Andronov93,Berdiyorov09}. We point out that the critical current $j_c$ decreases considerably with
increasing width of the sample, while $j_{cd}$ only moderately decreases (see dashed curve in Fig. \ref{fig2} for
$w=160$ nm). The weakly-dissipative state is still characterized by vortex nucleation near the corner (see panel 4),
but starts at low current (low $j_c$) and occurs in a broader range of the applied current ($j_c<j<j_{cd}$).

\begin{figure} [t]
\includegraphics[width=\linewidth]{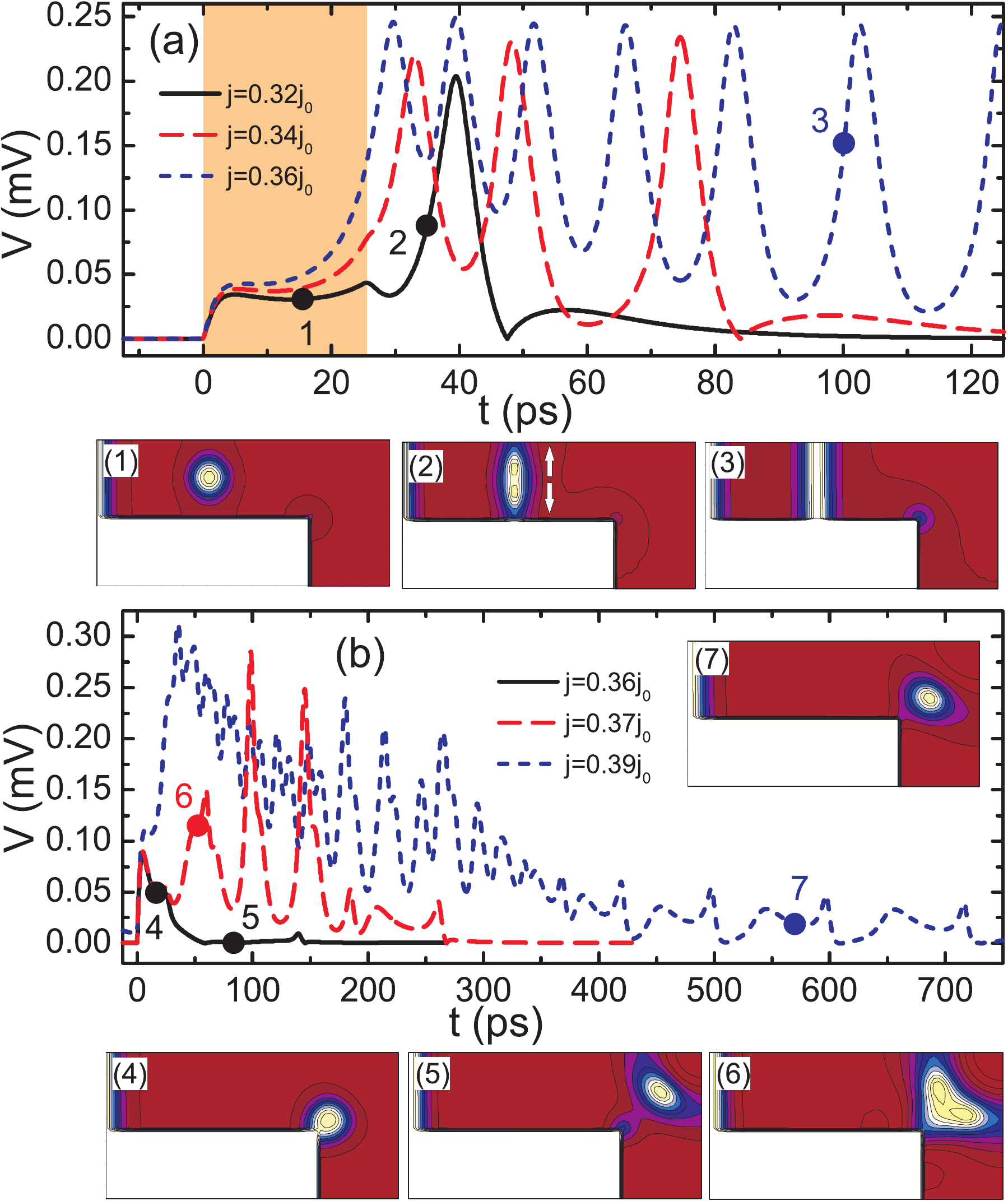}
\caption{\label{fig3}(Color online) Voltage vs. time characteristics of the sample (of size $L$=400 nm and $w$=110 nm)
at $T=0.9T_c$ for different applied currents. A single-photon (with pulse duration $\Delta t=25$ ps (shaded area in
(a)) and spot radius $R=15$ nm) acts: (a) in the middle of the stripe, c.f. panel 1, or (b) near the inner corner, c.f.
panel 4. Panels 1-7 show snapshots of $|\psi|^2$ (zoomed at the part of interest) at times indicated on the $V(t)$
curves. Arrows in panel 2 show the direction of motion of the unbinding vortex-antivortex pair.}
\end{figure}

In what follows, we study the response of our system to a single-photon absorption event. Following the effective
temperature approach \cite{Giazotto}, we assume that the single photon creates a hotspot of radius $R$, where the local
temperature becomes $T=2$ $T_c$ (see Ref. \cite{zotova} for the effect of such instant increase of the temperature).
First, we consider the case when the photon acts in the middle of the sample away from the corner. Fig. \ref{fig3}(a)
shows the $V(t)$ characteristics of the sample with $L=400$ nm and $w=110$ nm for different values of the applied
current density $j$. For each value of $j$, we started from the state obtained during the current increasing regime
(i.e. states from Fig. \ref{fig2}). The photon acts on the sample over the time interval of $\Delta t=25$ ps (the
shaded area in Fig. \ref{fig3}(a)), creating a hotspot of radius $R=15$ nm (c.f. panel 1 in Fig. \ref{fig3}). At low
bias currents ($j<0.315j_0$) the superconducting state is stable against the photon action (not shown here). With
increasing applied current, the system reacts to the photon absorption event by nucleating a vortex-antivortex pair
(see panel 2), which is subsequently unbound and split towards the edges of the sample by the current, resulting in a
voltage pulse (see solid curve in Fig. \ref{fig3}(a)). The system relaxes back into its initial state after the vortex
and the antivortex have left the sample and no voltage signal is observed at later times. The amplitude and duration of
the voltage signal, as well as the delay in the response of the system to the photon, all depend on the applied current
value. At larger current (still below $j_c$), several vortex-antivortex pairs can be formed one after another leading
to several voltage pulses per single photon absorption (see dashed curve in Fig. \ref{fig3}(a)). With further
increasing $j$, the fast moving vortices create a normal belt across the sample (panel 3) bringing the sample into the
resistive state. A finite average voltage develops across this resistive region which now can be detected
electronically (see dotted curve in Fig. \ref{fig3}(a)). In this sense, for applied currents just below $j_c$, we
confirm the recently predicted {\it vortex-assisted single-photon counting mechanism} \cite{bula,zotova}, where each
photon is detected thanks to the generated periodic motion of kinematic vortex-antivortex pairs \cite{Berdiyorov09}.

Fig. \ref{fig3}(b) shows the voltage response of the system when the photon acts near the inner corner of the sample
(c.f. panel 4). The photoresponse of the sample becomes totally different from the case discussed in Fig.
\ref{fig3}(a): (i) the photon generates a vortex near the inner corner of the sample (panel 5) at applied current
values much smaller than the critical one (i.e., starting from $j=0.25j_0$). (ii) The amplitude/duration of the voltage
signal due to the motion of this vortex is much smaller/longer as compared to the signal due to the vortex-antivortex
unbinding (compare solid curves in Figs. \ref{fig3}(a) and (b)), which indicates the slow motion of the vortex. (iii)
Instabilities can occur during the photon absorption (see dashed curve) due to the formation and dissociation of
multiquanta vortices (panel 6) near the inner corner, where supercurrent density is maximal (see panel 1$'$ in Fig.
\ref{fig2}). (iv) No transition into the phase-slip state (similar to the one in inset 3 of Fig. \ref{fig2}) occurs and
the resistive state is characterized by a weakly dissipative (``cold'') vortex crossing \cite{bula} (see the dotted
curve and inset 7 in Fig. \ref{fig3}(b)). Therefore, although the photon acting near the corner does trigger the
nucleation of superconducting vortices at currents much smaller than $j_c$, it {\it does not bring the system to a
higher resistive state} that can be detected electronically.

\begin{figure} [t]
\includegraphics[scale=0.45]{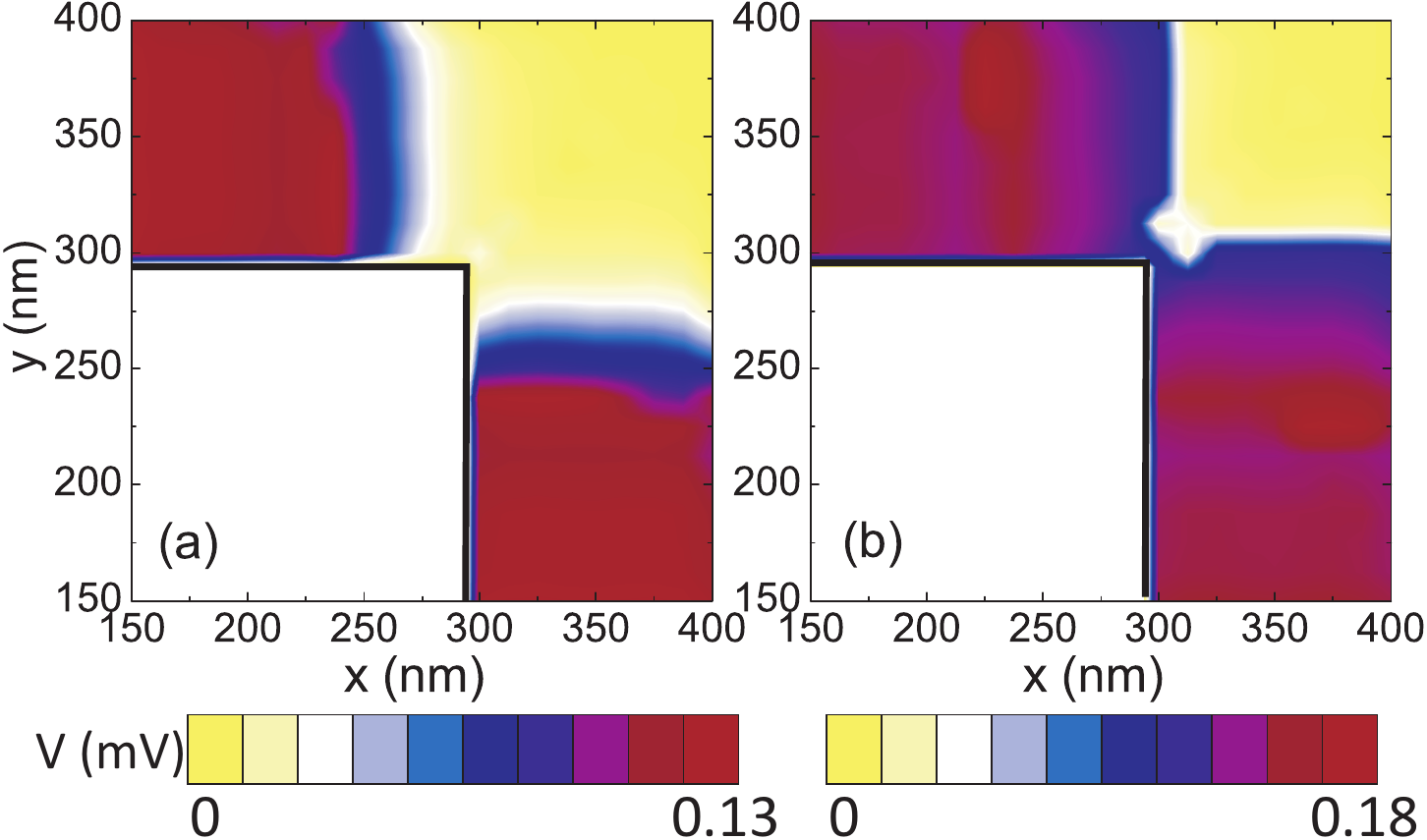}
\caption{\label{fig4}(Color online) Phase diagram: voltage response of the sample of Fig. \ref{fig3} as a function of
the location of the photon absorption for two values of the applied current: (a) $j=0.373j_0$ and (b) $j=0.383j_0$.}
\end{figure}

Our findings are summarized in Fig. \ref{fig4} where we plot a phase diagram, showing the voltage response of the
sample as a function of the location of the photon absorption. The voltage is averaged over the time interval $t=0.4-2$
ns to exclude the oscillating features of transient voltages during the photon action (such as those shown in Fig.
\ref{fig3}). It is clear from this figure that {\it the photoresponse of the meandering detector is spatially
dependent} with maximal sensitivity away from the turning point. Only at currents above $j_c$ (Fig. \ref{fig4}(b)) the
regions near the turning corner of the sample become responsive to the incident photon. Here we assume that the small
and short-lived resistance due to the slow moving vortices nucleated near the inner corner of the sample (see solid
curve in Figs. \ref{fig2} and \ref{fig3}(b)) is not sufficient to induce current at the shunting load, so that it is
beyond the sensitivity of the photodetecting measurement.

In conclusion, we confirmed that {\it vortex-assisted energy dissipation} is the dominant origin for single photon
counting in superconducting single-photon detectors. However, sensitivity of the commonly used meandering detectors is
{\it spatially dependent} with maximal response to single-photon absorption away from the sharp corners, and with
small response at the turns. Our suggestion is therefore to perform measurements at currents just under $j_c$, the
critical current which by itself induces a moving vortex at the inner corners of the meander, and calculate the photon
density as the number of detected counts over the area of only the straight parts of the detector. Alternatively, one
can perform measurements at currents just under $j_{cd}$, which corresponds to the jump to large dissipation. In this
weakly dissipative regime every impact of a photon would lead to a count, but one must cope with enduring heating
issues and non-zero dynamic resistance of the detector in the electronic circuit. However, the latter strategy may be
the only one suitable for wider meandering detectors, which provide larger surface to capture photons, but suffer from
particularly low $j_c$.

This work was supported by the Flemish Science Foundation (FWO-Vl).
G.R.B. acknowledges individual support from FWO-Vl.

\end{document}